\documentclass[11pt]{article}
\usepackage{amsmath}
\usepackage{amsfonts}
\usepackage{amssymb}

\input{tcilatex}

\begin{document}

\title{\textbf{On a Schwarzschild like metric}}
\author{\textbf{M. Anastasiei and I. Gottlieb}}
\date{}
\maketitle

\vspace{1cm}

\begin{center}
\textit{Dedicated to the 85th birthday of Professor Cleopatra Mociu\c tchi}
\end{center}

\vspace{1cm}

\textbf{Abstract.} In this short Note we would like to bring into the
attention of people working in General Relativity a Schwarzschild like
metric found by Professor Cleopatra Mociu\c tchi in sixties. It was obtained
by the A. Sommerfeld reasoning from his treatise "Elektrodynamik" but using
instead of the energy conserving law from the classical Physics, the
relativistic energy conserving law.

\section{ From Newton's Mechanics to General Relativity}

To begin with, let's recall the laws of Newton's mechanics :

\begin{enumerate}
\item A particle moves with constant velocity if no force acts on it.

\item The acceleration of a particle is proportional to the force acting on
it.

\item The forces of action and reaction are equal and opposite.
\end{enumerate}

These laws hold in their simplest form only in inertial frames. The fact is
stated as the Galilean principle of relativity :

{\it Galilean PR : The laws of mechanics have the same form
in all inertial frames.}

It implies the classical velocity addition law that has many confirmations
in classical mechanics but does not hold for the propagation of light. The
speed of light $c$ is the same in all inertial frame. Thus the PR was
extended in the form of the next two postulates :

{\it P1 : Physical laws have the same form in all inertial frames.}

{\it P2 : The speed of light is finite and equal in all inertial frames.}

These are basic for Special Relativity (SR). The SR is due to A.Einstein who
came at the conclusion that the concepts of space and time are \textit{%
relative} i.e. dependent on the reference frame of an observer.
Mathematically, the postulates P1 and P2 implies that two inertial frames
are connected by a Lorentz transformation of coordinates. This fact leads to
a new law for the addition of velocities and to two real physical effects :
length contraction and time dilatation. The Lorentz transformations show
also that is more natural to speak about \textit{spacetime} as a set of
points (events) with four coordinates than about space and time separately.
The squared differential distance between neighboring events is given by the
Lorentz invariant quadratic form
\begin{equation}
ds^2 = c^2dt^2 -dx^2-dy^2-dz^2.  \tag{1.1}
\end{equation}
However, the non-inertial (accelerated) frames do exist. A. Einstein noticed
that the dynamical effects of a gravitational field and an accelerated frame
can not be distinguished and he formulated the principle of equivalence (PE).

{\it $PE_1$: Every non-inertial frame is locally equivalent to some
gravitational field.}

This is more completely restated as

{\it $PE_2$: The gravitational field is locally equivalent to the field
of inertial forces of a convenient accelerated frame of reference.}

A freely falling reference frame is called a locally inertial frame. Using
this term the PE can be also restated as

{\it $PE_3$: At every point in an arbitrary gravitational field we can
choose a locally inertial frame in which the laws of Physics take the same
form as in SR.}

A. Einstein considered SR as incomplete because of the role played by the
inertial frames. He generalized the relativity of inertial motions to the
relativity of all motions by formulating the general principle of relativity.

{\it General PR : The form of physical laws is the same in all reference
frames.}

Mathematically, the general PR can be realized with the help of the
principle of general covariance.

{\it General covariance : The form of physical laws does not depend on
the choice of coordinates.}

This principle says that the physical laws have to be given in tensorial
form. For details and historical motivations of the principles reviewed in
the above we refer to the first chapter of the book [1]

\section{The Einstein equations}

Let's relabel the time and the spatial variable in a spacetime $M4$ as $%
x=(x^{0}=t,x^{1},x^{2},x^{3})$ . Then (1.1) is a particular form of the
following (the Einstein 's convention on summation is implied)
\begin{equation}
ds^{2}=g_{ij}dx^{i}dx^{j},i,j,k,...=0,1,2,3.  \tag{2.1}
\end{equation}

Here the entries of the matrix $g_{ij}(x)$ are the local components of a
pseudo or semi- Riemannian metric $g$ on $M_4$. We suppose that the
canonical form of the matrix $g_{ij}(x)$ is diag$(+,-,-,-)$. One says that $%
M_4$ is a Lorentz manifold. There exists an unique linear connection $\nabla
$ with the local coefficients given by the usual form of the Christofell
symbols which has no torsion and makes $g$ covariant constant i.e. $g_{ij;k}
=0$, where $;$ means the covariant derivative. Then the Ricci tensor is $%
R_{ik} = R^j_{i jk}$ where $R^h_{i jk} $ is the curvature tensor of $\nabla$
and the scalar curvature is $R= g^{ij}R_{ij}$. A. Einstein postulates the
basic equation for gravitational processes as it follows (the Einstein
equations)

\begin{equation}
EE: R_{ik} - \dfrac{1}{2}Rg_{ik} =\kappa T_{ik},  \tag{2.2}
\end{equation}
where $\kappa$ is a constant and $T_{ik}$ is the energy -momentum tensor.
The tensor $T_{ik}$ refers to the free formations (of the substance having
non-zero rest mass, the free electromagnetic field, etc. This should be
divergence free since the Einstein tensor (given by the left hand of the
Einstein equation (2.2)) is divergence free.

Alternatively, the EE can be derived from Hamilton's principle
\begin{equation}
\delta\int (bR +L)\sqrt{-g}d\tau =0,  \tag{2.3}
\end{equation}
where $\sqrt{-g}d\tau$ is the four dimensional volume element and in the
Hilbert- Palatini Lagrange function $bR +L$, $b$ is a constant, whereas the
function $L$ leads to $T_{ik}$.

Using the Einstein equations one may try to determine $g_{ik}$ assuming that
$T_{ik}$ is given or to take a special form of $g_{ik}$ and to determine $%
T_{ik}$. The second task is easier. The first is much harder. In absence of
the matter (void or empty space) we have $T_{ik}=0$ and a contraction with $%
g^{ik}$ in (2.2) leads to a simpler form of the EE : $R_{ik}= 0$. The
simplest and the most important exact solution of this last equation is the
Schwarzschild metric to be discussed in the next Section. For details we refer to [2].

\section{ The Schwarzschild metric}

The Schwarzschild metric is the first exact solution of the equation
\begin{equation}
R_{ik} =0.  \tag{3.1}
\end{equation}

It was found by Karl Schwarzschild in 1916 and played an important role in
confirming the predictions of the GR theory. Now there are many ways to
deduce it. It is valid in the empty space surrounding a static
body with mass spherical symmetry.
Passing to spherical coordinates $(x^{0},x^{1},x^{2},x3)=(t,r,\theta
,\varphi )$ , the spherical symmetry and an obvious re-scaling of $r$
reduces $ds^{2}$ to the following form (see [2,p.45])
\begin{equation}
ds^{2}=Adt^{2}-Bdr^{2}-r^{2}(d\theta ^{2}+sin^{2}\theta \varphi ^{2},
\tag{3.1}
\end{equation}

where because of the chosen signature we must have $%
A>0$ and $B>0$ and the both tend to $1$ at large distance from the source ( $%
r\rightarrow \infty $ ). Then one computes the Christofell symbols $\Gamma
_{jk}^{i}$ which are inserted in a formula for $R_{ik}$ ,( 5.27 in [2]). The Einstein equations (3.1) give first that the product $AB$
is a constant (equal to $1$ at $r$ tends to $\infty $). Hence $B=1/A$. Then $%
(r/B)^{\prime }=1$, where ' denotes the derivative with respect to $r$ and
upon integration one gets the Schwarzschild metric (SM)
\begin{equation}
ds^{2}=(1-\dfrac{\lambda }{r})dt^{2}-(1-\dfrac{\lambda }{r}%
)^{-1}dr^{2}-r^{2}(d\theta ^{2}+sin^{2}\theta d\varphi ^{2}),\lambda =\dfrac{%
2GM}{c^{2}},  \tag{3.2}
\end{equation}%
where $G$ is the universal gravitational constant and $M$ is the mass of the
central body (star, planet...). We stress that this metric is valid in the
empty space outside of the central body.

In his treatise [3], A. Sommerfeld derives the SM directly from the
equivalence principle in form $PE_3$. Here is his reasoning.

Let be a Point $P$ lying in the empty space surrounding a central mass $M$ which is distributed with spherical symmetry, its center being a point $O$. Within spherical coordinate $P$ will be determined by the $r, \theta, \varphi$. Let us consider a local frame of reference in $P$, one of the axes of the frame being $OP$. A second frame of reference, with the coordinates $\dot{r}, \theta. \varphi$ slides on the $OP$ along the first, the relative velocity being $v$, that is the velocity a material point of mass $m$ has under the action of the gravitational field of the central body. The metric connected to the mass $m$ will be the Minkowskian one, that is
\begin{equation}
ds^{2}=c^2d\dot{t}^{2}-
d\dot{r}^{2}-\dot{r}^{2}(d\theta ^{2}+sin^{2}\theta d\varphi ^{2}),\tag{3.3}
\end{equation}
An observer having a fixed position in $P$ will notice a momentary contraction of the distance as well as a momentary dilatation of the duration, that is
\begin{equation}
dr = \sqrt{1- v^2/c^2}d\dot{r}^2; dt = d\dot{t}/ \sqrt{1-v^2/c^2}, \tag {3.4}
\end{equation}
and so (3.3) becomes
\begin{equation}
ds^{2}=c^2(1-v^2/c^2)dt^{2}-
(1-v^2/c^2)^{-1} dr^{2}-r^{2}(d\theta ^{2}+sin^{2}\theta  d\varphi ^{2}),\tag{3.5}
\end{equation}
where $r=\dot{r}$, as we are talking about the same point $P$. It remains to determine the velocity $v=v(r)$, where the dependence on $r$ is due to the fact that we a talking about an accelerated material point, therefore $v$ varies permanently during the motion. In order to do this, A. Sommerfeld has used the energy conserving law from the classical Physics:
\begin{equation}
\dfrac{1}{2}mv^2 +m V(r) = const., \tag {3.6}
\end{equation}
where $ V(r) = \dfrac{GM}{r}$ is the gravitational field potential of the central mass.
When $r\rightarrow \infty$ the constant is zero and so
\begin{equation}
\dfrac{v^2}{c^2}=\dfrac {\lambda}{r} ,  \lambda = 2GM/c^2. \tag {3.7}
\end{equation}
Using this in (3.4) one finds the SM in the form (3.2).

The main point was here the use of the energy conserving law from (3.6). But accordingly to the $PE$ in its from $PE_3$ it is not only more natural but it is just compulsory the use of the relativistic law of energy conservation as a law from $SR$. It is true that the $SR$ includes also the classical laws but only in limits and with some nuances due to different groups of symmetries. This remark belongs to Professor Cleopatra Mociu\c tchi. Based on it she uses the relativistic law of energy conservation and so she arrived to and studied in the sixties, [4-8], a Schwarzschild like metric as follows.

The relativistic law of energy conservation has the form
\begin{equation}
(m-m_0)c^2 - \dfrac{GmM}{r}=0,    m=m_0\sqrt{1-v^2/c^2}. \tag {3.8}
\end{equation}
It comes out that
\begin{equation}
\sqrt{1-v^2/c^2}=1 - \mu /r. \tag {3.9}
\end{equation}
an so (3.5) reduces to what we call the  Mociu\c tchi metric  (MM) :
\begin{equation}
ds^{2}=c^2(1-\dfrac{\mu }{r})^2dt^{2}-(1-\dfrac{\mu }{r})^{-2}
dr^{2}-r^{2}(d\theta ^{2}+sin^{2}\theta d\varphi ^{2}),\mu =\dfrac{
GM}{c^{2}}= \lambda/2,  \tag{3.10}
\end{equation}

Here some properties of the Mociu\c tchi metric :
\begin{enumerate}
\item It returns all the General Relativity tests,
\item It reduces to the  SM  if $(\mu /r)^2 \approx 0$,
\item The black hole radius that results from it is half from the Schwarzschild radius,
\item When passing through the radius of the black hole, the variables $r$ and $t$ do not change significance between themselves in the case of MM, contrary to SM.
\item The MM results from (2.3), $L$ is the Lagrange function of the gravitational field of the central mass.
\end{enumerate}

The last enumerated property may be connected with the remark that if one computes the Einstein tensor $ R_{ik} - \dfrac{1}{2}Rg_{ik}$ for the MM it comes out that four of its components are non null. In the other words the MM is not an exact solution of the Einstein equation with the energy-momentum tensor $T_{ij} =0$. However we derived it in the hypothesis of the absence of  the matter and fields (in the empty space outside of a star or a planet). The contradiction could be eliminated if we assume that even in this empty space a kind of manifestation of the gravitational field does exist. But we can do this if we slightly modify the PE as follows :

{\it $P'E$: The {\bf only dynamical effects } of a gravitational field are  locally equivalent to the field
of inertial forces of a convenient accelerated frame of reference.}

Moreover, some results on interaction of various fields confirm , [9-11], a more general principle of equivalence :

{\it $P"E$: The {\bf dynamical effects of certain fields } are  locally equivalent to the field
of inertial forces of a convenient accelerated frame of reference.}

\bigskip

Author's adresses: \\
\begin{flushleft}
Mihai Anastasiei\\
\begin{tabular}{llr}
Faculty of Mathematics & \qquad and & \quad Mathematical Institute "O.Mayer"\\
``Al.I. Cuza'' University & \qquad & Romanian Academy, \\
Bd. Carol I, no. 11, & \qquad & Ia\c si, ROMANIA \\
700506 Ia\c{s}i, ROMANIA
\end{tabular}
E-mail: anastas@uaic.ro\\
\bigskip
Ioan Gottlieb,\\
Faculty of Physics\\
"Al.I.Cuza" University \\
Bd. Carol I, no.11, \\
700506, Ia\c si, ROMANIA \\

E-mail : gottlieb@uaic.ro
\end{flushleft}

\end{document}